\documentclass[a4paper,11pt]{article}
\usepackage{pos}
\usepackage{subcaption}
\newcommand{\beq}{\begin{equation}}
\newcommand{\eeq}{\end{equation}}
\newcommand{\beqs}{\begin{eqnarray}}
\newcommand{\eeqs}{\end{eqnarray}}
\usepackage{graphicx}
\usepackage{dsfont}
\usepackage{dcolumn}
\usepackage{multirow}
\newcommand{\nn}{\nonumber}
\usepackage{array}
\newcolumntype{C}[1]{>{\centering\let\newline\\\arraybackslash\hspace{0pt}}m{#1}}
\newcolumntype{L}[1]{>{\raggedright\let\newline\\\arraybackslash\hspace{0pt}}m{#1}}

\title{\boldmath Chimera baryon spectrum in the Sp(4) completion of composite Higgs models}

\author*[a]{Ho Hsiao}
\author[b]{Ed Bennett}
\author[c,d]{Deog Ki Hong}
\author[e,d]{Jong-Wan Lee}
\author[a,f,g]{C.-J. David Lin}
\author[h,b]{Biagio Lucini}
\author[i]{Maurizio Piai}
\author[j]{Davide Vadacchino}

\affiliation[a]{Institute of Physics, National Yang Ming Chiao Tung University, 1001 Ta-Hsueh Road, Hsinchu 30010, Taiwan}
\affiliation[b]{Swansea Academy of Advanced Computing, Swansea University (Bay Campus), Fabian Way, SA1 8EN Swansea, Wales, United Kingdom}
\affiliation[c]{Department of Physics, Pusan National University, Busan 46241, Korea}
\affiliation[d]{Extreme Physics Institute, Pusan National University, Busan 46241, Korea}
\affiliation[e]{Particle Theory and Cosmology Group, Center for Theoretical Physics of the Universe,
Institute for Basic Science (IBS), Daejeon, 34126, Korea}
\affiliation[f]{Center for High Energy Physics, Chung-Yuan Christian University, Chung-Li 32023, Taiwan}
\affiliation[g]{Centre for Theoretical and Computational Physics, National Yang Ming Chiao Tung University, 1001 Ta-Hsueh Road, Hsinchu 30010, Taiwan}
\affiliation[h]{Department of Mathematics, Faculty of Science and Engineering, Swansea University (Bay Campus), Fabian Way, SA1 8EN Swansea, Wales, United Kingdom}
\affiliation[i]{Department of Physics, Faculty of Science and Engineering, 
Swansea University (Park Campus), Singleton Park, SA2 8PP Swansea, Wales, United Kingdom}
\affiliation[j]{Centre for Mathematical Science, University of Plymouth, Plymouth, PL4 8AA, United Kingdom}

\emailAdd{thepaulxiao.sc09@nycu.edu.tw}
\emailAdd{e.j.bennett@swansea.ac.uk}
\emailAdd{dkhong@pusan.ac.kr}
\emailAdd{j.w.lee@ibs.re.kr}
\emailAdd{dlin@nycu.edu.tw}
\emailAdd{b.lucini@swansea.ac.uk}
\emailAdd{m.piai@swansea.ac.uk}
\emailAdd{davide.vadacchino@plymouth.ac.uk}

\abstract{In strongly coupled gauge theories that serve as completions of composite Higgs models, the fermionic bound states formed by fermions (hyperquarks) transforming in different representations, called chimera baryons, could serve as top partners, by embedding of the Standard Model appropriately. We report our results on the spectrum of chimera baryons in the $Sp(4)$ gauge theory
with hyperquarks transforming in fundamental and two-index antisymmetric representations. For this study, we adopt the quenched approximation. We investigate the mass hierarchy  between the lightest chimera baryons with different quantum numbers, as a function of the lattice parameters. Inspired by baryon chiral effective field theory, and the Akaike Information Criterion,
we perform a first extrapolation to the continuum and massless-hyperquark limit.}

\FullConference{The 40th International Symposium on Lattice Field Theory (Lattice 2023)\\
July 31st - August 4th, 2023\\
Fermi National Accelerator Laboratory\\}

\begin{document}
\maketitle

\section{Introduction}

The Higgs boson discovery~\cite{ATLAS:2012yve,CMS:2012qbp} intensifies the quest for a deeper understanding of nature. While experimental evidence supports the standard model (SM), the triviality of its scalar sector suggests it is an effective field theory (EFT) with a finite ultraviolet (UV) cut-off. 
Composite Higgs models (CHMs) stand out as potential theories, as they provide a natural framework for accommodating a light Higgs boson~\cite{Dugan:1984hq,Panico:2015jxa,Cacciapaglia:2020kgq,
Ferretti:2013kya}.
These models introduce a novel strongly coupled sector with an asymptotically-free gauge theory coupled to fermions (hyperquarks), where the SM Higgs boson emerges as a pseudo-Nambu-Goldstone boson (PNGB) associated with a global symmetry of the new strong interaction. Furthermore, CHMs can tackle the flavor problem by incorporating partial compositeness for the top quark~\cite{Kaplan:1991dc}.
By coupling the theory to hyperquarks in two gauge group representations and embedding the SM gauge group appropriately, bound states formed by hyperquarks in different representations can share the quantum numbers as the top quark. These bound states, known as top partners, contribute to the mass of the top quark through mixing.

Our collaboration has been developing an extensive programme of lattice studies~\cite{Bennett:2017kga, Bennett:2019jzz,Bennett:2019cxd, Bennett:2020qtj, Bennett:2022yfa, Bennett:2022ftz, Bennett:2023wjw} focused on the $Sp(4)$ gauge theory coupled to $N_{f}=2$ Dirac fermions in the fundamental, $(f)$, and $n_{f}=3$  Dirac fermions in the two-index antisymmetric, $(as)$, representations of the gauge group~\cite{Barnard:2013zea}.
In the $(f)$ sector, the $SU(4)/Sp(4)$ coset, due to the pseudoreality~\cite{Peskin:1980gc}, provides candidates for the SM Higgs doublet with only one additional Goldstone mode. 
As the $(as)$ representation is real, the global symmetry  is $SU(6)$, broken to $SO(6)$~\cite{Peskin:1980gc}. The $SU(3)$ subgroup of the unbroken $SO(6)$ can be identified with the QCD gauge group~\cite{Barnard:2013zea,Ferretti:2013kya}.
In this framework, the top partners,  dubbed chimera baryons, are composed of two $(f)$ and one $(as)$ hyperquarks.

In this contribution, we report the spectrum of the low-lying chimera baryons, sourced by the following operators:
 \beqs
 {\mathcal{O}}_{\rho}^{ijk,5} &\equiv& Q^{i\,a}_{\alpha} (C\gamma^{5})_{\alpha\beta} Q^{j\,b}_{\beta} \Omega^{ad}\Omega^{bc} \Psi^{k\,cd}_{\rho} \, , \label{eq:chim_bar_src} \\ 
 {\mathcal{O}}_{\rho}^{ijk,\mu} &\equiv& Q^{i\,a}_{\alpha} (C\gamma^{\mu})_{\alpha\beta} Q^{j\,b}_{\,\beta} \Omega^{ad}\Omega^{bc} \Psi^{k\,cd}_{\rho} \, ,\label{eq:chim_bar_src_mu}
 \eeqs
 where $Q$ and $\Psi$ are $(f)$ and $(as)$ hyperquarks, respectively, $a,b,c,d$ are hypercolor indices, $\alpha, \beta$, $\rho$ are spinor indices, $i,j,k$ are flavor indices,
$\gamma^5$ and $\gamma^{\mu}$ are $4\times 4$ Dirac matrices, $C$ is the charge conjugation matrix and $\Omega$ is the symplectic matrix.
Given both operators in Eqs.~(\ref{eq:chim_bar_src}) and~(\ref{eq:chim_bar_src_mu}) overlap with even- and odd-parity states, and the operator ${\mathcal{O}}^{\mu}$ couples to both spin-1/2 and 3/2 states, we perform parity and spin projections to isolate the states with designed quantum numbers, see Section IIIA in Ref.~\cite{Bennett:2023mhh} for detailed discussions. The lightest state sourced by ${\mathcal{O}}^{5}$, denoted as $\Lambda_{\rm CB}$, is of spin-1/2 and even-parity.  In the case of ${\mathcal{O}}^{\mu}$, the lightest spin-1/2 even-parity state and spin-3/2 even-parity state are $\Sigma_{\rm CB}$ and $\Sigma^\ast_{\rm CB}$, respectively.
Both $\Lambda_{\rm CB}$ and $\Sigma_{\rm CB}$ are top-partner candidates.

Considering this is the first systematic lattice calculation for the chimera baryon mass spectrum in the $Sp(4)$ gauge theory, we perform simulations in the quenched approximation.
The standard Wilson plaquette is applied for gauge-field action, while we employ valence fermions described by the Wilson-Dirac lattice action in calculations of the hyperquark propagators. We generate five ensembles, listed in Tab.~\ref{tab:ENS}, with different lattice spacing allowing for the continuum extrapolation.
The gradient flow method is applied to set the scale~\cite{Luscher:2010iy}, and following the notation in Ref.~\cite{Bennett:2022ftz} we express the masses in unit of the gradient flow $w_0$~\cite{BMW:2012hcm}, denoting $\hat{m}=w_0m$. The square of the pseudoscalar meson mass is adopted as a reference scale for the hyperquark masses~\cite{Bennett:2019cxd}.  The masses of pseudoscalar mesons with $(f)$ and $(as)$ hyperquarks are denoted as $m_{\rm PS}$ and $m_{\rm ps}$, respectively.
Regrading the chimera baryons, $m_{\Lambda_{\rm CB}}$, $m_{\Sigma_{\rm CB}}$ and $m_{\Sigma^\ast_{\rm CB}}$ denote the mass of even-parity $\Lambda_{\rm CB}$, $\Sigma_{\rm CB}$, and $\Sigma^\ast_{\rm CB}$, respectively.

For a detailed discussion including lattice formulation, projections and analysis approaches, we refer readers to Ref.~\cite{Bennett:2023mhh} and references therein.

\begin{table}
\small
   \caption{
  Gauge ensembles generated for this study. We display the bare coupling $\beta$, the lattice size, $N_t\times N^3_s$, the average plaquette $\left < P \right >$, and the gradient-flow scale $w_0/a$. The gradient-flow scales are taken from Ref.~\cite{Bennett:2019cxd}.
   \label{tab:ENS}}
\begin{center}
    \begin{tabular}{| c | c | c | c | c | c|}
    \hline\hline
        Ensemble& $\beta$   & $N_t\times N^3_s$ & $\left < P \right >$  & $w_0/a$  \\ \hline
        QB1	    & $7.62$    & $48\times24^3$    & 0.6018905(94)               & 1.448(3)      \\
        QB2	    & $7.7$     & $60\times48^3$    & 0.6088003(35)             & 1.6070(19)    \\ 
        QB3	    & $7.85$    & $60\times48^3$    & 0.6203811(28)              & 1.944(3)      \\ 
        QB4	    & $8.0$     & $60\times48^3$    & 0.6307426(27)             & 2.3149(12)    \\
        QB5	    & $8.2$     & $60\times48^3$    & 0.6432300(25)              & 2.8812(21)    \\
        \hline
    \end{tabular}
   \end{center}
   \vspace{-2em}
\end{table}

\section{Mass hierarchy}
 
To study the hyperquark-mass dependence of the chimera-baryon mass hierarchy, we perform the calculations at various bare masses, see Appendix A in Ref.~\cite{Bennett:2023mhh}.
We observe a decreasing ratio $m_{\Lambda_{\rm CB}}/m_{\Sigma_{\rm CB}}$   with increasing $\hat{m}^2_{\rm ps}$. This ratio eventually approaches unity in the large-$\hat{m}^{2}_{\rm ps}$ regime. 
A similar pattern emerges with varying $\hat{m}_{\rm PS}^{2}$ except that $m_{\Lambda_{\rm CB}}/m_{\Sigma_{\rm CB}}$ never approaches unity in the region of large $m_{\rm PS}^{2}$. When the $(as)$ hyperquark is heavy, the ratio shows a mild dependence on $\hat{m}_{\rm PS}^{2}$ and is primarily influenced by $\hat{m}^{2}_{\rm ps}$.
Throughout our entire range of hyperquark masses, $\Lambda_{\rm CB}$ is not lighter than $\Sigma_{\rm CB}$.
Likewise, we find that $\Sigma^\ast_{\rm CB}$ is consistently heavier than $\Sigma_{\rm CB}$ and $\Lambda_{\rm CB}$. The mass gap between them decreases with increasing $\hat{m}^{2}_{\rm PS}$ and $\hat{m}^{2}_{\rm ps}$. This behavior aligns with expectations from heavy-hyperquark spin symmetry~\cite{Isgur:1989vq}, as increasing hyperquark masses suppress the effects of spin that contribute to the mass difference between $\Sigma_{\rm CB}$ and $\Sigma^\ast_{\rm CB}$.

\section{Mass extrapolation}

We extrapolate the chimera baryon mass to the continuum and massless-hyperquark limit by drawing inspiration from baryon chiral perturbation theory in QCD \cite{Jenkins:1990jv, Bernard:1995dp}, and from its lattice realization~\cite{Beane:2003xv}.
Considering the following ansatz, we perform uncorrelated fits to the chimera-baryon masses using polynomial functions of $\hat{m}_{\mathrm{PS}}$, $\hat{m}_{\mathrm{ps}}$, and the lattice spacing, $\hat{a}$,
\beqs
\hat{m}_{\textrm{CB}} = \hat{m}_{\rm CB}^\chi &+& F_2 \hat{m}_{\rm PS}^2 + A_2 \hat{m}_{\rm ps}^2 + L_1 \hat{a}  \nn \\ 
&+& F_3 \hat{m}_{\rm PS}^3 + A_3 \hat{m}_{\rm ps}^3 + L_{2F} \hat{m}_{\rm PS}^2 \hat{a} + L_{2A}  \hat{m}_{\rm ps}^2 \hat{a} \nn \\
&+& F_4 \hat{m}_{\rm PS}^4 + A_4  \hat{m}_{\rm ps}^4 + C_{4}  \hat{m}_{\rm PS}^2  \hat{m}_{\rm ps}^2\,.
\label{eq:fitting_func}
\eeqs
Here ${\rm CB} = \Lambda_{\rm CB}$, $\Sigma_{\rm CB}$ or $\Sigma_{\rm CB}^\ast$, and $\hat{m}_{\rm CB}^\chi$ denotes the mass of the chimera baryon in the continuum and massless-hyperquark limit.
The coefficients $F_j$ and $A_j$ are the low energy constants (LECs) associated with corrections to $\hat{m}_{\rm CB}$ at the $j$-th power in $\hat{m}_{\rm PS}$  and $\hat{m}_{\rm ps}$, respectively, while the coefficient $C_{4}$ governs the cross-term proportional to $\hat{m}^{2}_{\mathrm{PS}}\hat{m}^{2}_{\mathrm{ps}}$.
The LECs $L_1$, $L_{2F}$, and $L_{2A}$, correspond to  corrections due to the finite lattice spacing, $\hat{a}$.
We only consider polynomial dependence of $\hat{m}_{\rm CB}$ on pseudoscalar meson masses, omitting possible logarithmic terms, and perform the fit independently for each chimera baryon.

Our first attempt, fitting to Eq.~(\ref{eq:fitting_func}) with the entire dataset, results in a large value of $\chi^2/N_{\rm d.o.f.}$---indicating a very poor fit.
As a result, we impose a set of cuts on pseudoscalar meson masses, ($\hat{m}_{\mathrm{PS, cut}}, \hat{m}_{\mathrm{ps, cut}} $), to fit a subset including data points that sit within the cuts.
The initial set of cuts is chosen as $(\hat{m}_{\mathrm{PS, cut}}, \hat{m}_{\mathrm{ps, cut}} )=(0.52, 0.52)$, which ensures the minimal subset contains $13$ data points.
We then increase the cut values, $\hat{m}_{\mathrm{PS, cut}}$ and $\hat{m}_{\mathrm{ps, cut}}$, independently, in steps of  $0.05$, and introduce the condition, $am_{\rm PS} < 1$ and $am_{\rm ps} < 1$, on the data points within that set of cuts. 
The maximum cut values stop at $(\hat{m}_{\mathrm{PS, cut}}, \hat{m}_{\mathrm{ps, cut}} )=(1.07, 1.87)$, where the whole dataset is included.
In this way, we construct $158$ distinct data sets. However, the $\chi^2/N_{\rm d.o.f.}$ values of fitting each data set to Eq.~(\ref{eq:fitting_func}) are still large.
Thus, we proceed the following fitting strategy.

The $158$ data sets are fitted to five fit ansatze based upon truncating Eq.~(\ref{eq:fitting_func}) to include a reduced number of fitting parameters.
We first consider the ansatz, dubbed M2, restricted to the first line of Eq.~(\ref{eq:fitting_func}), which contains 
 $\hat{m}^{\chi}_{\mathrm{CB}}$ and corrections quadratic in pseudoscalar-meson masses and linear in lattice spacing.
 The ansatz M3 is introduced by incorporating also corrections up to cubic in the pseudoscalar-meson masses, as well as the lattice-spacing corrections, $\hat{m}_{\textrm{PS}}^2 \hat{a}$ and $\hat{m}_{\textrm{ps}}^2 \hat{a}$. 
Finally, we incorporate the three highest-order  terms from Eq.~(\ref{eq:fitting_func}) individually. Within ansatze MF4, MA4, and MC4, these additions correspond to introducing $F_4 \hat{m}_{\textrm{PS}}^4$, $A_4 \hat{m}_{\textrm{ps}}^4$, or $C_{4} \hat{m}_{\textrm{PS}}^2 \hat{m}_{\textrm{ps}}^2$, respectively.
\begin{table}
\small
    \caption{List of the terms in Eq.~(\ref{eq:fitting_func}) that are associated to each fit ansatz used in our analysis. \label{tab:ansatz} }
    \vspace{-1em}
\begin{center}
    \begin{tabular}{|c|c|c|c|c|c|c|c|c|c|c|c|}
    \hline
         Ansatz & $\hat{m}_{\rm CB}^\chi$ & $\hat{m}_{\rm PS}^2$ & $\hat{m}_{\rm ps}^2$ & $\hat{m}_{\rm PS}^3$ & $\hat{m}_{\rm ps}^3$ & $\hat{m}_{\rm PS}^4$ & $\hat{m}_{\rm ps}^4$ & $\hat{m}_{\rm PS}^2 \hat{m}_{\rm ps}^2$ & $\hat{a}$ &  $\hat{m}_{\rm PS}^2 \hat{a}$ &  $\hat{m}_{\rm ps}^2 \hat{a}$ \\ \hline
         M2  & \checkmark & \checkmark & \checkmark & - & - & - & - & - & \checkmark & - & -\\ \hline
         M3  & \checkmark & \checkmark & \checkmark & \checkmark & \checkmark & - & - & - & \checkmark & \checkmark & \checkmark\\ \hline
         MF4 & \checkmark & \checkmark & \checkmark & \checkmark & \checkmark & \checkmark & - & - & \checkmark  & \checkmark & \checkmark \\ \hline
         MA4 & \checkmark & \checkmark & \checkmark & \checkmark & \checkmark & - & \checkmark & - & \checkmark  & \checkmark & \checkmark \\ \hline
         MC4 & \checkmark & \checkmark & \checkmark & \checkmark & \checkmark & - & - & \checkmark & \checkmark & \checkmark & \checkmark \\
         \hline
    \end{tabular} 
\end{center}
\end{table}

With these five ansatze, tabulated with their associated terms in Tab.~\ref{tab:ansatz}, and $158$ data sets, we obtain $790$ analysis procedures. In order to quantitatively select the best procedure, we compute the Akaike information criterion (AIC)~\cite{Akaike:1998zah,Jay:2020jkz}. For each analysis procedure we compute 
\beq\label{eq:AIC}
{\rm AIC} \equiv \chi^2 + 2k + 2N_{\rm cut}\,,
\eeq
where $\chi^2$ represents the standard chi-square, $k$ is the number of fit parameters, and $N_{\rm cut}$ is the number of data points removed by the cuts, $(\hat{m}_{\mathrm{PS, cut}}, \hat{m}_{\mathrm{ps, cut}} )$.
The corresponding probability weight is defined as
\beq\label{eq:AIC_W}
W \equiv \frac{1}{\mathcal{N}}\exp\left [ {-\frac{1}{2} {\rm AIC} } \right ]\,,
\eeq
where the normalization factor $\mathcal{N}$ assures the sum of $W$ over all $790$ analysis procedures equals to one.  
\begin{table}
\small
\caption{Low-energy constants in Eq.~(\ref{eq:fitting_func}) for each chimera baryon, as determined by the best analysis procedure in the ansatz at $(\hat{m}_{\mathrm{PS, cut}}, \hat{m}_{\mathrm{ps, cut}} )$. The missing coefficients are set to zero.}
\begin{center}
    \begin{tabular}{| c | c c c c  c  c  |}
    \hline\hline
    CB  	&$\hat{m}_{\rm CB}^\chi $	     & $ F_2  $ 	&  $A_2 $ 	   & $F_3 $  &   $A_3 $ 	 & $C_4$ \\ \hline

$\Lambda_{\rm CB}$ & 0.999(27) & 0.709(61) & 0.383(10) & -0.150(34) & -0.092(4)  & -0.026(5) \\

$\Sigma_{\rm CB}$  & 0.841(21) & 0.815(77) & 0.558(13) & -0.252(63) & -0.161(7)  & -0.078(7) \\

$\Sigma^\ast_{\rm CB}$  & 1.259(34) & 0.360(110) & 0.393(29) & -0.071(93) & -0.129(16)  & - \\
\hline
 CB  &  $L_1 $ 	   & $L_{2F} $	  &   $L_{2A}$ & Ansatz   &  $\hat{m}_{\rm PS,cut}$ &  $\hat{m}_{\rm ps,cut}$ \\ \hline

$\Lambda_{\rm CB}$   & -0.137(43) & 0.071(69) & 0.004(11) &MC4  & 1.07 & 1.87 \\

$\Sigma_{\rm CB}$   & -0.141(33) & 0.187(65) & -0.021(16) & MC4& 0.77 & 1.47     \\

$\Sigma^\ast_{\rm CB}$  & -0.335(54) & 0.342(83) & 0.004(30) & M3~ ~  & 0.82 & 1.17  \\
\hline
\end{tabular}
   \end{center}
	\vspace{-1em}
   \label{tab:best_fit}
\end{table}

By applying this fitting strategy to the three chimera baryons independently, we find that the optimal choice of analysis procedure for $\Lambda_{\rm CB}$ is MC4 at $(\hat{m}_{\mathrm{PS, cut}}, \hat{m}_{\mathrm{ps, cut}} )=(1.07, 1.87)$. 
Regrading $\Sigma_{\rm CB}$ and $\Sigma^\ast_{\rm CB}$, the best analysis procedures are MC4 at $(\hat{m}_{\mathrm{PS, cut}}, \hat{m}_{\mathrm{ps, cut}} )=(0.77, 1.47)$ and M3 at $(\hat{m}_{\mathrm{PS, cut}}, \hat{m}_{\mathrm{ps, cut}} )=(0.82, 1.17)$, respectively.
We present corresponding LECs obtained by the best analysis procedures in Tab.~\ref{tab:best_fit}, see more details in Section IIIC of Ref.~\cite{Bennett:2023mhh}.

\begin{figure}
	\centering
	\includegraphics[width=0.95\textwidth]{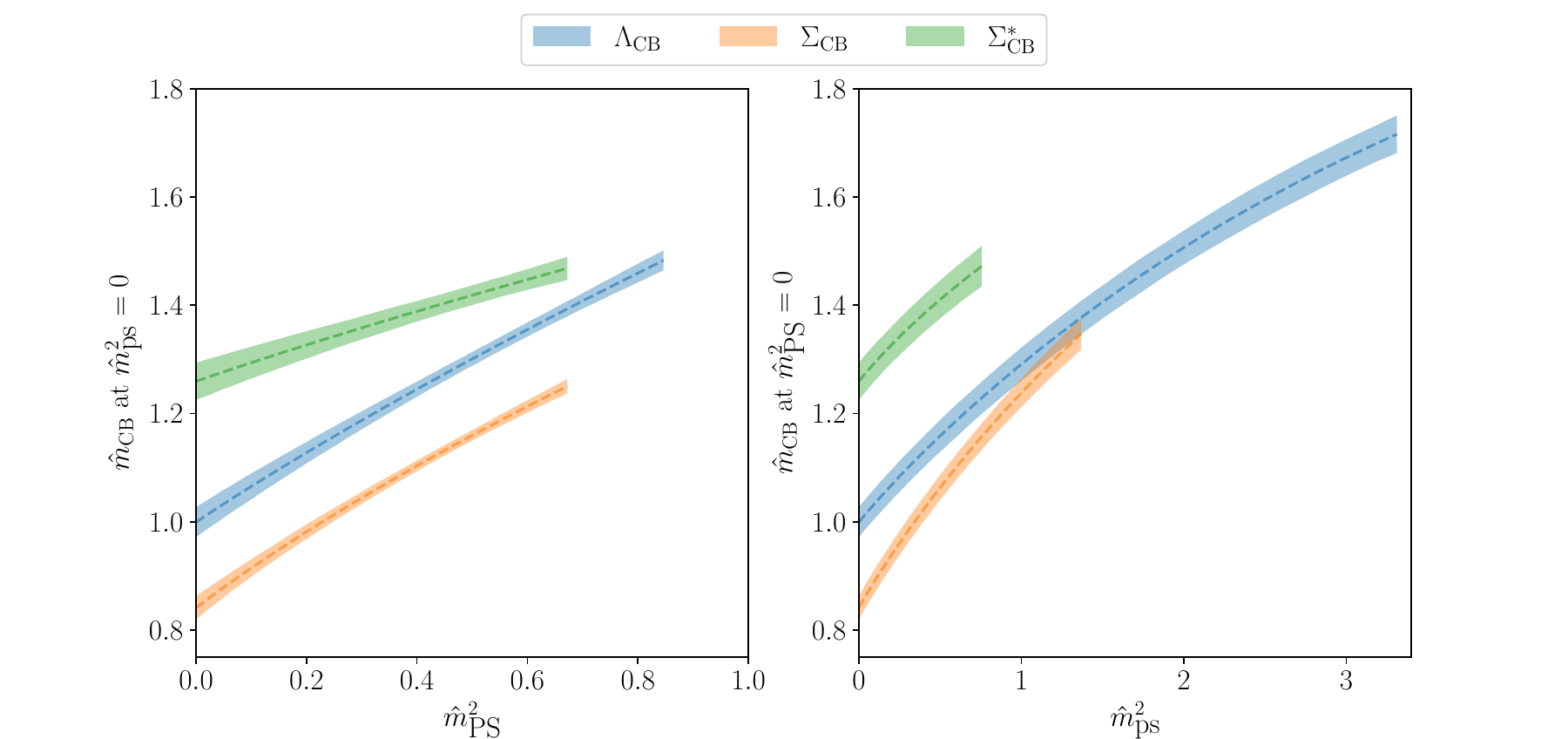}
	\caption{
 Mass of chimera baryons ${\Lambda_{\rm CB}}$, ${\Sigma_{\rm CB}}$, and ${\Sigma^{\ast}_{\rm CB}}$ as a function of $\hat{m}^{2}_{\rm PS}$ (left) and $\hat{m}^{2}_{\rm ps}$ (right) in the limit of vanishing lattice spacing. 
The plots are generated with $\hat{m}^{2}_{\rm ps} = 0$ (left) and $\hat{m}^{2}_{\rm PS}=0$ (right), based on best-fit LECs from Tab.~\ref{tab:best_fit}, with bands representing statistical errors.  
 Bands extend horizontally between zero and optimal choices of $\hat{m}^{2}_{\rm PS, cut}$ (left) and $\hat{m}^{2}_{\rm ps, cut}$ (right).}
	\label{fig:m_massless}
\end{figure}

Having determined the LECs, we explore the dependence of the chimera-baryon masses in the continuum limit by taking $\hat{a}=0$ in Eq.~(\ref{eq:fitting_func}).
In Fig.~\ref{fig:m_massless}, the left (right) panel shows the change of $\hat{m}_{\Lambda_{\rm CB}}$, $\hat{m}_{\Sigma_{\rm CB}}$, and $\hat{m}_{\Sigma_{\rm CB}}$ with respect to the variation of  $\hat{m}_{\rm PS}$ ($\hat{m}_{\rm ps}$) in the limit where $\hat{m}_{\rm ps} = 0$ ($\hat{m}_{\rm PS} = 0$).
The mass hierarchy, $ \hat{m}_{\Sigma_{\rm CB}} \lesssim \hat{m}_{\Lambda_{\rm CB}} < \hat{m}_{\Sigma^{\ast}_{\rm CB}}$, is observed across the entire range of hyperquark masses explored in this study.
Specifically, the masses $\hat{m}_{\Lambda_{\rm CB}}$ and $\hat{m}_{\Sigma_{\rm CB}}$ are compatible only in the regime where $(as)$ hyperquarks are substantially heavy. This hierarchy carries non-trivial implications for the development of composite Higgs models with top partial compositeness.

\begin{figure}[b]
	\centering
	\includegraphics[width=0.95\textwidth]{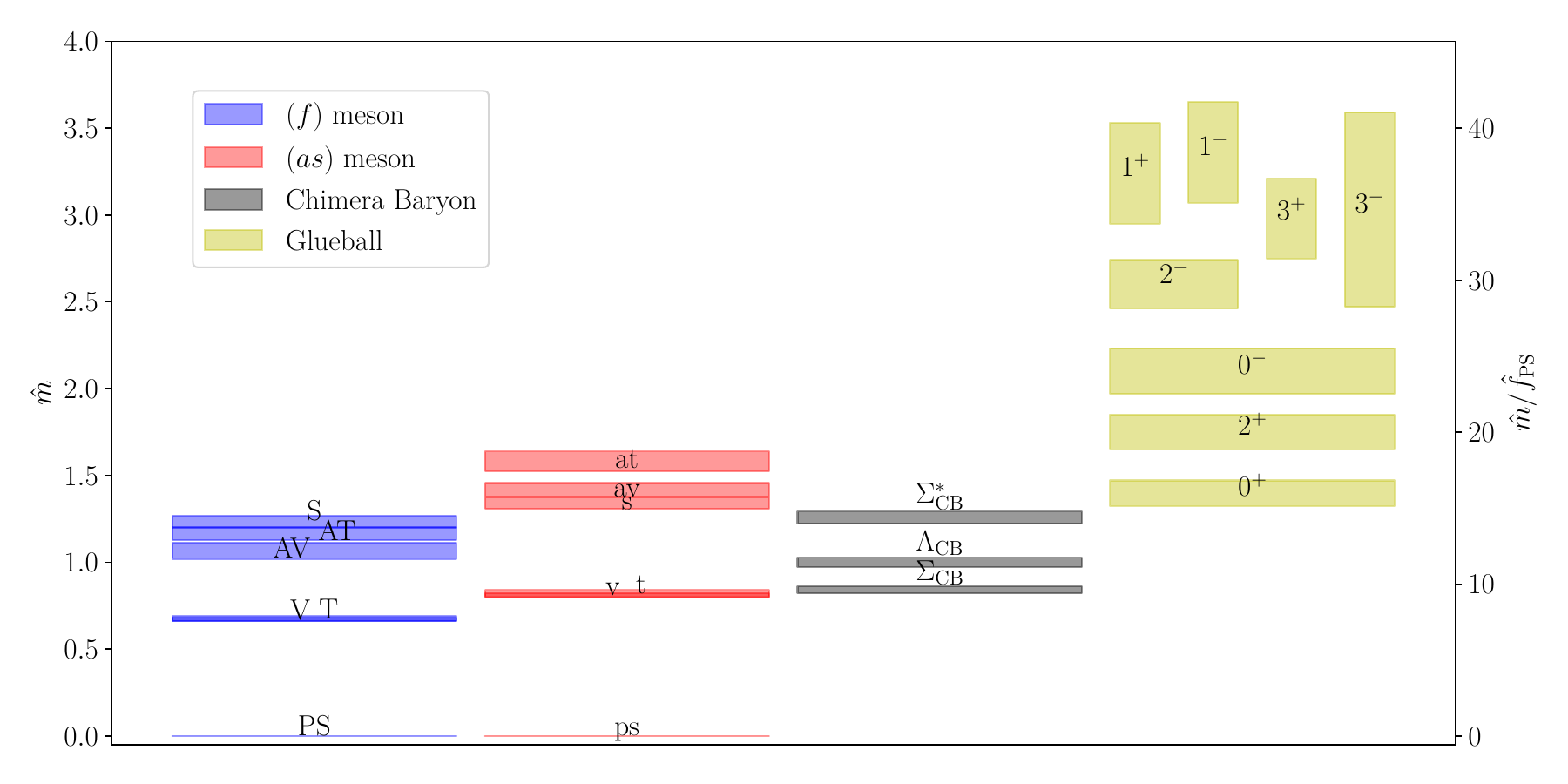}
	\caption{Quenched spectrum of the $Sp(4)$ gauge theory in the continuum and massless-hyperquark limit. The pseudoscalar, vector, tensor, axial-vector, axial-tensor and scalar mesons composed of fundamental (antisymmetric)  hyperquarks are denoted as PS (ps), V (v), T(t), AV (av), AT (at) and S (s), respectively, while glueball states are labelled by $J^{P}$. 
 The results of mesons and glueballs are taken from our previous works in Refs.~\cite{Bennett:2019cxd, Bennett:2020qtj}.
    }
	\label{fig:quench_spec}
	\vspace{-2em}
\end{figure}

In addition, we extend the plot of the mass spectrum by including mesons and glueballs in the theory.
In Fig.~\ref{fig:quench_spec}, meson and glueball masses are taken from our previous studies in the quenched approximation~\cite{Bennett:2019cxd, Bennett:2020qtj}, while the chimera-baryon masses are taken in the massless-hyperquark limit ($\hat{m}^{2}_{\rm ps} = \hat{m}^{2}_{\rm PS}=0$).
Mesons labeled with capital letters consist of $(f)$ hyperquarks, whereas those represented by lowercase letters are made of $(as)$ hyperquarks.
All masses have been extrapolated to the continuum and massless-hyperquark limit. They are displayed in both gradient-flow units (vertical axis on the left-hand side) and units of the fundamental pseudoscalar meson decay constant (vertical axis on the right-hand side).
The masses of the top-partner candidates, $\Lambda_{\rm CB}$ and $\Sigma_{\rm CB}$, closely align with those of the $(as)$ vector mesons.

\section{Summary and Outlook}

We presented our measurement, performed in the quenched approximation and in the Sp(4) gauge theory, of the mass spectrum of chimera baryons, $\Lambda_{\rm CB}$, $\Sigma_{\rm CB}$, and $\Sigma^\ast_{\rm CB}$. The first two such states are top partner candidates in a class of composite Higgs model with top partial compositeness. We examined their mass hierarchy and performed continuum and massless-hyperquark extrapolations. This study serves as a foundation for future lattice simulations with dynamical hyperquarks. 

\begin{acknowledgments}

{\footnotesize
The work of HH and CJDL is supported by the Taiwanese MoST grant 109-2112-M-009-006-MY3 and NSTC grant 112-2112-M-A49-021-MY3.
The work of EB has been supported by the UKRI Science and Technology Facilities Council (STFC)
Research Software Engineering Fellowship EP/V052489/1, and by the ExaTEPP project EP/X017168/1.
The work of DKH was supported by Basic Science Research Program through the National Research Foundation of Korea (NRF) funded by the Ministry of Education (NRF-2017R1D1A1B06033701).
The work of JWL was supported in part by the National Research Foundation of Korea (NRF) grant funded 
by the Korea government(MSIT) (NRF-2018R1C1B3001379) and by IBS under the project code, IBS-R018-D1. 
The work of DKH and JWL was further supported by the National Research Foundation of Korea (NRF) grant funded by the Korea government (MSIT) (2021R1A4A5031460).
The work of BL and MP has been supported in part by the STFC 
Consolidated Grants  No. ST/P00055X/1, ST/T000813/1, and ST/X000648/1.
 BL and MP received funding from
the European Research Council (ERC) under the European
Union’s Horizon 2020 research and innovation program
under Grant Agreement No.~813942. 
The work of BL is further supported in part 
by the Royal Society Wolfson Research Merit Award 
WM170010 and by the Leverhulme Trust Research Fellowship No. RF-2020-4619.
The work of DV is supported by STFC under Consolidated
Grant No.~ST/X000680/1.
Numerical simulations have been performed on the 
Swansea University SUNBIRD cluster (part of the Supercomputing Wales project) and AccelerateAI A100 GPU system,
on the local HPC
clusters in Pusan National University (PNU) and in National Yang Ming Chiao Tung University (NYCU),
and on the DiRAC Data Intensive service at Leicester.
The Swansea University SUNBIRD system and AccelerateAI are part funded by the European Regional Development Fund (ERDF) via Welsh Government.
The DiRAC Data Intensive service at Leicester is operated by 
the University of Leicester IT Services, which forms part of 
the STFC DiRAC HPC Facility (www.dirac.ac.uk). The DiRAC 
Data Intensive service equipment at Leicester was funded 
by BEIS capital funding via STFC capital grants ST/K000373/1 
and ST/R002363/1 and STFC DiRAC Operations grant ST/R001014/1. 
DiRAC is part of the National e-Infrastructure.
\newline

{\bf Open Access Statement}---For the purpose of open access, the authors have applied a Creative Commons 
Attribution (CC BY) licence  to any Author Accepted Manuscript version arising.

{\bf Research Data Access Statement}---The results for Sp(4) are based on preliminary analysis.
Further analysis and the data generated for this manuscript will be released together with an upcoming publication~\cite{Bennett:2023mhh}.
Alternatively, data and code can be obtained from the authors upon request.

} 

\end{acknowledgments}

\bibliographystyle{JHEP}
{\footnotesize \bibliography{refs}}

\end{document}